\documentclass[pra,twocolumn]{revtex4}

\usepackage{graphicx}
\usepackage{bm}
\usepackage{amsmath}
\usepackage[psamsfonts]{amssymb}
\usepackage{type1cm}
\usepackage{pst-all}
\newcommand{\futo}[1]{{\bm #1}}
\newcommand{\fx}{\futo{x}}
\newcommand{\fy}{\futo{y}}
\newcommand{\tani}{{\openone}}

\newcommand{\ket}[1]{{\left\vert{#1}\right\rangle}}
\newcommand{\ketrei}{{\ket{0}}}
\newcommand{\ketichi}{{\ket{1}}}
\newcommand{\gyouretsu}[1]{\begin{pmatrix}#1\end{pmatrix}}

\newcommand{\cu}{controlled-$U$}
\newcommand{\diag}{\mathrm{diag}}
\newcommand{\Shita}[2]{{\Theta_{#1}(#2)}}
\newcommand{\kyo}{{i}}

\newcommand{\fukuso}{\mathbb{C}}
\newcommand{\kochi}[2]{{\sigma(#1,#2)}}
\newcommand{\kochiab}{\kochi{a}{b}}

\newcommand{\sacc}[1]{\begin{subarray}{c}#1\end{subarray}}
\def\yokotsuki#1#2{\setbox0\hbox{$#1$}
  \setbox2\null \ht2\ht0 \dp2\dp0 \box2^#2 \! \box0}
\newcommand{\tenchi}[1]{{\yokotsuki{#1}{t}}}
\newcommand{\hako}[1]{\makebox[2em]{\rule{0mm}{1.6ex}#1}}
\newcommand{\kairohabamoto}{1.6}
\newcommand{\kairotatemoto}{3}
\newcommand{\kairohaba}{16}
\newcommand{\kairotate}{30}
\newcommand{\kairohabahan}{8}
\newcommand{\kairohabap}{17.5}
\newcommand{\kairokaeshi}{-1.25}
\newcommand{\kairotsunagihabamoto}{.5}
\newcommand{\kairotsunagihaba}{7}
\newcommand{\kairohabajimoto}{.8}
\newcommand{\senichi}{5}
\newcommand{\nakaichi}{16}
\newcommand{\nakaichige}{14}
\newcommand{\senni}{25}

\newcommand{\kairocu}[2]{
  \begin{pspicture}(0,0)(\kairohabamoto,\kairotatemoto)\psset{unit=1mm}
    \pnode(\kairokaeshi,\senichi){a}
    \rput(\kairohabahan,\senichi){\ovalnode{b}{\hako{$#2$}}}
    \pnode(\kairohabap,\senichi){c}
    \pnode(\kairokaeshi,\senni){e}
    \rput(\kairohabahan,\senni){\dianode{f}{$#1$}}
    \pnode(\kairohabap,\senni){g}
    \ncline{a}{b}
    \ncline{b}{c}
    \ncline{e}{f}
    \ncline{f}{g}
    \ncline{b}{f}
  \end{pspicture}
}
\newcommand{\kairocuni}[2]{
  \begin{pspicture}(0,0)(\kairohabamoto,\kairotatemoto)\psset{unit=1mm}
    \pnode(\kairokaeshi,\senichi){a}
    \rput(\kairohabahan,\senichi){\dianode{b}{$#1$}}
    \pnode(\kairohabap,\senichi){c}
    \pnode(\kairokaeshi,\senni){e}
    \rput(\kairohabahan,\senni){\ovalnode{f}{\hako{$#2$}}}
    \pnode(\kairohabap,\senni){g}
    \ncline{a}{b}
    \ncline{b}{c}
    \ncline{e}{f}
    \ncline{f}{g}
    \ncline{b}{f}
  \end{pspicture}
}
\newcommand{\kairoichi}[1]{
  \begin{pspicture}(0,0)(\kairohabamoto,\kairotatemoto)\psset{unit=1mm}
    \pnode(\kairokaeshi,\senichi){a}
    \rput(\kairohabahan,\senichi){\ovalnode{b}{\hako{$#1$}}}
    \pnode(\kairohabap,\senichi){c}
    \pnode(\kairokaeshi,\senni){d}
    \pnode(\kairohabap,\senni){e}
    \ncline{a}{b}
    \ncline{b}{c}
    \ncline{d}{e}
  \end{pspicture}
}
\newcommand{\kaironi}[1]{
  \begin{pspicture}(0,0)(\kairohabamoto,\kairotatemoto)\psset{unit=1mm}
    \pnode(\kairokaeshi,\senichi){a}
    \pnode(\kairohabap,\senichi){b}
    \pnode(\kairokaeshi,\senni){c}
    \rput(\kairohabahan,\senni){\ovalnode{d}{\hako{$#1$}}}
    \pnode(\kairohabap,\senni){e}
    \ncline{a}{b}
    \ncline{c}{d}
    \ncline{d}{e}
  \end{pspicture}
}
\newcommand{\kairotsunagi}{
  \begin{pspicture}(0,0)(\kairotsunagihabamoto,\kairotatemoto)\psset{unit=1mm}
    \pnode(\kairokaeshi,\senichi){a}
    \pnode(\kairotsunagihaba,\senichi){b}
    \pnode(\kairokaeshi,\senni){c}
    \pnode(\kairotsunagihaba,\senni){d}
    \ncline{a}{b}
    \ncline{c}{d}
  \end{pspicture}
}
\newcommand{\kairotsunagiio}[4]{
  \begin{pspicture}(0,0)(\kairotsunagihabamoto,\kairotatemoto)\psset{unit=1mm}
    \pnode(\kairokaeshi,\senichi){a}
    \pnode(\kairotsunagihaba,\senichi){b}
    \pnode(\kairokaeshi,\senni){c}
    \pnode(\kairotsunagihaba,\senni){d}
    \ncline{a}{b}\naput[npos=0]{#1}\naput[npos=1]{#2}
    \ncline{c}{d}\naput[npos=0]{#3}\naput[npos=1]{#4}
  \end{pspicture}
}
\newcommand{\kairoten}{
  \begin{pspicture}(0,0)(\kairotsunagihabamoto,\kairotatemoto)\psset{unit=1mm}
    \pnode(\kairokaeshi,\senichi){a}
    \pnode(\kairotsunagihaba,\senichi){b}
    \pnode(\kairokaeshi,\senni){c}
    \pnode(\kairotsunagihaba,\senni){d}
    \ncline[linestyle=dotted]{a}{b}
    \ncline[linestyle=dotted]{c}{d}
  \end{pspicture}
}
\newcommand{\kairoji}[1]{
  \begin{pspicture}(0,0)(\kairohabajimoto,\kairotatemoto)\psset{unit=1mm}
    \rput(5,\nakaichi){#1}
  \end{pspicture}
}
\newcommand{\kairohako}[1]{
  \begin{pspicture}(0,0)(\kairohabamoto,\kairotatemoto)\psset{unit=1mm}
    \pnode(\kairokaeshi,\senichi){a}
    \pnode(0,\senichi){b}
    \psframe(0,0)(\kairohaba,\kairotate)
    \pnode(\kairohaba,\senichi){c}
    \pnode(\kairohabap,\senichi){d}
    \pnode(\kairokaeshi,\senni){e}
    \pnode(0,\senni){f}
    \pnode(\kairohaba,\senni){g}
    \pnode(\kairohabap,\senni){h}
    \ncline{a}{b}
    \ncline{c}{d}
    \ncline{e}{f}
    \ncline{g}{h}
    \rput(\kairohabahan,\nakaichige){$#1$}
  \end{pspicture}
}

\begin{document}

\title{The Controlled-$U$ and Unitary Transformation in Two-Qudit}
\author{FUNAHASHI Kunio}
\email{funahasi@isc.meiji.ac.jp}
\affiliation{ Division of Natural Science, Izumi Campus, Meiji University, 
  Tokyo 168-8555, Japan}
\begin{abstract}
  We concretely construct an extension of the \cu gate in qudit
  from some elementary gates.
  We also construct unitary transformation in two-qudit by means of 
  the extended \cu gate and show the universality of it.
\end{abstract}

\maketitle

\section{Introduction}
\label{sec:introduction}

In study on networks of quantum computer, qubit is mainly used
for unit of circuits.
A qubit has two states, $\ketrei$ and $\ketichi$, and preserves
their superposition.
It is a unit of computation.
A network of quantum computer consists of a bundle of them
and quantum computation is a sequence of quantum gates,
which are unitary transformations, on them.
Although we do not know what kind of unitary transformation is required,
it is shown that any unitary transformation can be constructed in
qubit\cite{deutsch,divincenzo}.

In qubit number of steps of computation tends to be large
because algorithm is given in two-value.
When network is constructed in actual physical system, there arises
decoherence which is caused by interaction with environment\cite{zurek,frasca}.
Also in this point of view, it is better that number of steps is small.
Further physical systems are not always two-level.
For example, atom has infinite energy levels.
There is no need to restrict to only two of them.

It is considered that unit of computation has three or more states.
This unit is called by ``qudit''.
Quantum computation in qudit is a sequence of unitary transformations
on a bundle of qudits as well as in qubit case.

Computation in qudit has some advantage to that in qubit.
Use of multi-valued unit may decrease number of steps.
This is favorable for the decoherence problem.
Waste of high excited states in physical systems may fall off.

As stated above, quantum gates in qudit are unitary transformations
on a bundle of qudits.
Such gates have been devised by some researchers\cite{muthukrishnan,
brylinski,karimipour,fujii1,fujii2,fujii3,fujii4,daboul}.
However, as far as we know, there is no work on construction of
unitary transformation from elementary gates concretely
and proof of universality of it.
It seems that the universality is approved on the analogy of qubit,
however, it should be proved strictly.
So in this paper we construct unitary transformation from elementary
gates concretely and show the universality of it.

The constitution of the paper is as follows.
In Sec.\ref{sec:elementarygates} we give some elementary gates.
We build any network by sequence of them.
We extend the controlled-unitary gate to qudit, which was introduced in
\cite{9nin} to construct unitary transformation in qubit in 
Sec.~\ref{sec:constructioncugate}.
In Sec.~\ref{sec:constr-unit-transf} we show that we can construct any
unitary transformation in two-qudit using the 
controlled-unitary gates constructed in the previous section.
The last section is devoted to the discussion.

\section{Elementary Gates for Qudit}
\label{sec:elementarygates}

There are some ways to construct networks in qudit and there may be
different elementary gates in each way.
So first we introduce elementary gates in our method.

We assume the following.
First we can perform any unitary transformation in a single qudit.
In spite of this fact there exist some important gates.
We introduce them in Sec.\ref{sec:elem-gate-single}.
Second there are at least one gate to connect two (or more) qudits
for entanglement.
We introduce such a gate in Sec.~\ref{sec:controlled-gate}.

We consider a $d$-level system, thus there are $d$ states in a qudit.
We label them as
\begin{equation*}
  \ket{k}\qquad(k=0,1,2,\ldots,d-1).
\end{equation*}
When we express states and operators in matrix form we identify it 
with $d$-dimensional vector
\begin{equation*}
  \ket{k}={}^t(0,0,\ldots,0,\overset{\sacc{k\\\lor}}{1},0,\ldots,0),
\end{equation*}
where $t$ means transpose.

\subsection{Elementary gates in a single qudit}
\label{sec:elem-gate-single}

%% As stated above, we assume that any unitary transformation is 
%% available in a single qudit.
%% However, there are some important gates.
%% We introduce two kinds of them.

First we introduce $P_{ab}$ which exchange two states whose
operation is
%\begin{equation*}
\begin{multline*}
  %\label{eq:exchangepab}
  P_{ab}\ket{c}=
  \left\{
    \begin{array}{@{\,}ll}
      \ket{b},&\text{if }\ket{c}=\ket{a}\\
      \ket{a},&\text{if }\ket{c}=\ket{b}\\
      \ket{c},&\text{otherwise}
    \end{array}
  \right.\\(a,b,c=0,1,2,\ldots,d-1).
%\end{equation*}
\end{multline*}
At this point $P_{ba}$ is the same with $P_{ab}$,
so we restrict to $a\leq b$ unessentially.
The diagram is
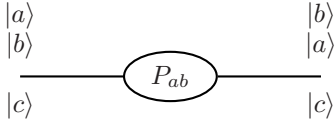
\begin{center}
  \begin{pspicture}(0,0)(4,4)\psset{unit=1mm}
    \pnode(0,20){a}
    \rput(20,20){\ovalnode{b}{\hako{$P_{ab}$}}}
    \pnode(40,20){c}
    \ncline{a}{b}\naput[npos=0]{$\begin{array}{c}\ket{a}\\\ket{b}\end{array}$}%
    \nbput[npos=0]{$\ket{c}$}
    \ncline{b}{c}\naput[npos=1]{$\begin{array}{c}\ket{b}\\\ket{a}\end{array}$}%
    \nbput[npos=1]{$\ket{c}$}
  \end{pspicture}
\end{center}
and its matrix form is
\begin{equation*}
%  \label{eq:pabmatirx}
  (P_{ab})_{ij}=\delta_{ij}+\delta_{ia}(-\delta_{ja}+\delta_{jb})+\delta_{ib}(\delta_{ja}-\delta_{jb})
\end{equation*}
which is one of elementary matrices.

$P_{ab}$ satisfies
\begin{equation*}
%  \label{eq:2}
  P_{ab}=\tenchi{P_{ab}}=P_{ab}^\dagger,\quad P_{ab}^2=\tani_d,
\end{equation*}
where $\dagger$ means hermitian conjugate.

Number of $P_{ab}$'s is ${}_dC_2=d(d-1)/2$, however, only $d-1$ of them
are more fundamental.
In fact if we put
\begin{equation*}
%  \label{eq:3}
  P_a\equiv P_{0a}\quad(a=1,2,\ldots,d-1),
\end{equation*}
others are expressed by
\begin{equation*}
%  \label{eq:4}
  P_{ab}=P_aP_bP_a.
\end{equation*}
Nevertheless in actual physical systems all gates may be handled equally.

Second we introduce another important gate which is an extension of the
Walsh-Hadamard gate.
The operation is
\begin{equation*}
  H_{ab}\ket{c}=
  \left\{
    \begin{array}{@{\,}rl}
      \dfrac{1}{\sqrt{2}}(\ket{a}+\ket{b}),&\text{if }\ket{c}=\ket{a}\\
      \dfrac{1}{\sqrt{2}}(\ket{a}-\ket{b}),&\text{if }\ket{c}=\ket{b}\\
      \ket{c},&\text{otherwise}
    \end{array}
  \right.
\end{equation*}
and its matrix form is
%\begin{equation*}
\begin{multline*}
%  \label{eq:walshhadamard}
  (H_{ab})_{ij}=\delta_{ij}
  +\delta_{ia}
  \left\{
    (-1+\frac{1}{\sqrt{2}})\delta_{ja}+\frac{1}{\sqrt{2}}\delta_{jb}
  \right\}\\
  +\delta_{ib}
  \left\{
    \frac{1}{\sqrt{2}}\delta_{ja}-(1+\frac{1}{\sqrt{2}})\delta_{jb}
  \right\}.
\end{multline*}
%\end{equation*}

$H_{ab}$ satisfies
\begin{equation*}
   H=\tenchi{H}=H^\dagger,\quad H^2=\tani_d.
\end{equation*}

Similar to $P_{ab}$, number of $H_{ab}$ is ${}_dC_2$, however, only one
gate is fundamental.
For example, we choose
\begin{equation*}
   H\equiv H_{01},
\end{equation*}
others are expressed by
\begin{equation*}
   H_{ab}=P_{0a}P_{1b}HP_{1b}P_{0a},\quad(1\leq a\leq b).
\end{equation*}

\subsection{Controlled gate}
\label{sec:controlled-gate}

%% There should exist at least one gate to connect two or more qudits because
%% if one does not exist, networks are decomposed to tensor product of qudits.

Now we introduce a symbol $\tilde C^a(U)$ which means the controlled gate.
$C$ means the controlled operation and exponent $a$ indicates the state of
the control bit in which the unitary transformation $U$ is applied.
The tilde over $C$ means two qudit operation.
In equation the operation of $\tilde C^a(U)$ is
\begin{equation*}
%  \label{eq:14}
  \tilde C^a(U)\ket{c}\ket{d}=
  \left\{
    \begin{array}{@{\,}rl}
      \ket{a}(U\ket{d}),&\text{if }\ket{c}=\ket{a}\\[5pt]
      \ket{c}\ket{d},&\text{otherwise}
    \end{array}
  \right..
\end{equation*}
Also we introduce the diagram of $\tilde C^a(U)$ as follows:
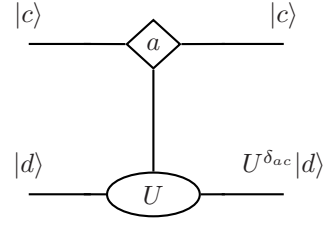
\begin{center}
  \kairotsunagiio{$\ket{d}$}{}{$\ket{c}$}{}\kairocu{a}{U}%
  \kairotsunagiio{}{$U^{\delta_{ac}}\ket{d}$}{}{$\ket{c}$}
\end{center}

As the controlled gate, the controlled-NOT or the controlled-$\sigma_z$
is usually used in qubit.
In this paper we extend the controlled-$\sigma_z$ in qudit.
There are some extensions.
For example, one of them is that if the control bit is some state
then all of the states in the target bit are shifted at once.
However our extension is the following.
\begin{quotation}
  If the control bit is $\ket{a}$, then the sign of $\ket{b}$ is
  reversed, otherwise the target bit is left alone.
\end{quotation}
In equation
%\begin{equation}
\begin{multline}
  \label{eq:8}
   \tilde C^a(M_b)\ket{c}\ket{d}=
  \left\{
    \begin{array}{@{\,}rl}
      -\ket{a}\ket{b},&\text{if }\ket{c}\ket{d}=\ket{a}\ket{b}\\[5pt]
      \ket{c}\ket{d},&\text{otherwise}
    \end{array}
  \right.,\\(a,b,c,d=0,1,2,\ldots,d-1),
\end{multline}
%\end{equation}
where $M_b$ is a single qudit operation which reverses the sign of $\ket{b}$:
\begin{equation*}
%  \label{eq:15}
  M_b\ket{c}=
  \left\{
    \begin{array}{@{\,}rl}
      -\ket{b},&\text{if }\ket{c}=\ket{b}\\[5pt]
      \ket{c},&\text{otherwise}
    \end{array}
  \right.\quad,(b,c=0,1,\ldots,d-1).
\end{equation*}
The matrix form of $\tilde C^a(M_b)$ is
%\begin{equation}
\begin{multline}
    \label{eq:9}
  (\tilde C^a(M_b))_{ij}=\delta_{ij}(1-2\delta_{i,ad+b}),\\(i,j=0,1,\ldots,d^2-1),
%\end{equation}
\end{multline}
where
\begin{equation*}
  M_b=\diag(1,1,\ldots,\overset{\sacc{b\\\lor}}{-1},\ldots,1),\quad(i,j=0,1,\ldots,d-1),
\end{equation*}
and the diagram is
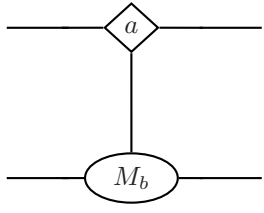
\begin{center}
%\kairoji{$\tilde M_b^a\equiv$}
\kairotsunagi\kairocu{a}{M_b}\kairotsunagi
\end{center}
From ~(\ref{eq:8})or ~(\ref{eq:9}), $\tilde C^a(M_b)$ apparently satisfies
\begin{equation*}
%  \label{eq:12}
  \tilde C^a(M_b)(\tilde C^a(M_b))^\dagger=\tani_{d^2}
\end{equation*}
and also
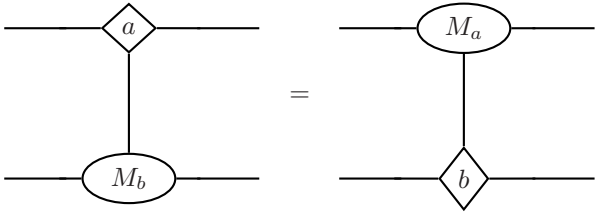
\begin{center}
  \kairotsunagi\kairocu{a}{M_b}\kairotsunagi
  \kairoji{$=$}
  \kairotsunagi\kairocuni{b}{M_a}\kairotsunagi
\end{center}
holds.

Number of $\tilde C^a(M_b)$'s is $d^2$, however, only one is fundamental.
For example, if we choose $\tilde C^{\overline a}(M_{\overline b})$, others
are obtained by
\begin{equation*}
%  \label{eq:10}
  \tilde C^a(M_b)=(\tani\otimes P_{a\bar a})(\tani\otimes P_{b\bar b})
  \tilde C^{\bar a}(M_{\bar b})
  (\tani\otimes P_{b\bar b})(\tani\otimes P_{a\bar a}).
\end{equation*}
Nevertheless, similar to $P_{ab}$, all $\tilde C^a(M_b)$'s may be
realized equally in actual physical systems.

\section{Construction of the Controlled-$U$ Gate}
\label{sec:constructioncugate}

\subsection{Preparation}
\label{sec:preparation}

In this section we construct the controlled-$U$ gate in qudit
by connecting the elementary gates given in the previous section.
Although we can consider some extensions, we adopt the following.
\begin{quotation}
  If the control bit is $\ket{a}$, then a unitary transformation $U$ is
  applied to the target bit, otherwise the target bit is left alone.
\end{quotation}
In equation
\begin{equation*}
%  \label{eq:13}
  \tilde C^a(U)\ket{c}\ket{b}=
\left\{
  \begin{array}{@{\,}rl}
    \ket{a}(U\ket{b}),&\text{if }\ket{c}=\ket{a}\\[5pt]
    \ket{c}\ket{b},&\text{otherwise}
  \end{array}
\right.,
\end{equation*}
and the diagram is
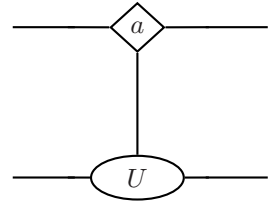
\begin{center}
\kairotsunagi\kairocu{a}{U}\kairotsunagi
\end{center}

First we construct the controlled-exchange gate $\tilde C^a(P_{bc})$
whose operation is
\begin{quotation}
  if the control bit is $\ket{a}$, then exchange the states $\ket{b}$
  and $\ket{c}$ in the target bit.
\end{quotation}
In equation
\begin{equation*}
%  \label{eq:16}
  \tilde C^a(P_{bc})\ket{d}\ket{e}=
  \left\{
    \setlength{\arraycolsep}{1pt}
    \begin{array}{@{\,}ll}
      \ket{a}\ket{c},&\quad\text{if }\ket{d}\ket{e}=\ket{a}\ket{b}\\[5pt]
      \ket{a}\ket{b},&\quad\text{if }\ket{d}\ket{e}=\ket{a}\ket{c}\\[5pt]
      \ket{d}\ket{e},&\quad\text{otherwise}
    \end{array}
  \right..
\end{equation*}
This gate is built by
\begin{equation*}
%  \label{eq:17}
  \tilde C^a(P_{bc})=(\tani\otimes H_{bc})\tilde C^a(M_c)(\tani\otimes H_{bc})
\end{equation*}
whose diagram is
\begin{widetext}
  \begin{center}
    \kairotsunagi\kairocu{a}{P_{bc}}\kairotsunagi
    \kairoji{$\equiv$}
    \kairotsunagi\kairoichi{H_{bc}}\kairocu{a}{M_c}\kairoichi{H_{bc}}\kairotsunagi
  \end{center}
\end{widetext}
Indeed, if the control bit is $\ket{a}$
\begin{equation*}
%  \label{eq:18}
  \tilde C^a(P_{bc})(\ket{a}\sum_{k=0}^{d-1}\alpha_k\ket{k})
  =\ket{a}(\alpha_c\ket{b}+\alpha_b\ket{c}+{\sum_{k=0}^{d-1}}^{\lor b,c}\alpha_k\ket{k})
\end{equation*}
and otherwise
\begin{equation*}
%  \label{eq:19}
  \tilde C^a(P_{bc})(\ket{l}\sum_{k=0}^{d-1}\alpha_k\ket{k})
  =(\tani\otimes\tani)\ket{l}\sum_{k=0}^{d-1}\alpha_k\ket{k}
  =\ket{l}\sum_{k=0}^{d-1}\alpha_k\ket{k}
\end{equation*}
where ${\sum}^{\lor b,c}$ means a sum except for $b$, $c$.

Second by means of $\tilde C^a(P_{bc})$ we construct the gate
whose operation is
\begin{quotation}
  if the control bit is $\ket{a}$, then perform the phase shift $e^{-\kyo\theta}$
  to $\ketrei$ and $e^{\kyo\theta}$ to $\ket{b}$.
\end{quotation}
In equation
\begin{equation*}
  \tilde C^a(\Theta_b(\theta))\ket{c}\ket{d}=
  \left\{
    \setlength{\arraycolsep}{1pt}
    \begin{array}{@{\,}rl}
      e^{-\kyo\theta}\ket{a}\ket{0},
      &\quad\text{if }\ket{c}\ket{d}=\ket{a}\ketrei\\[5pt]
      e^{\kyo\theta}\ket{a}\ket{b},
      &\quad\text{if }\ket{c}\ket{d}=\ket{a}\ket{b}\\[5pt]
      \ket{c}\ket{d},&\quad\text{otherwise}
    \end{array}
  \right.,
\end{equation*}
where $\Shita{b}{\theta}$ is a single qudit operation whose matrix form is
\begin{equation*}
%  \label{eq:21}
  \Shita{b}{\theta}\equiv\diag(\overset{\sacc{0\\\lor}}{e^{-\kyo\theta}},1,\ldots,
  \overset{\sacc{b\\\lor}}{e^{\kyo\theta}},\ldots,1).
\end{equation*}
This gate is realized by
%\begin{equation*}
\begin{multline*}
%  \label{eq:20}
  \tilde C^a(\Shita{b}{\theta})\equiv(\tani\otimes\Shita{b}{\frac{\theta}{4}})\tilde C^a(P_{0b})
  (\tani\otimes\Shita{b}{-\frac{\theta}{2}})\\\otimes \tilde C^a(P_{0b})
  (\tani\otimes\Shita{b}{\frac{\theta}{4}})
%\end{equation*}
\end{multline*}
and the diagram is
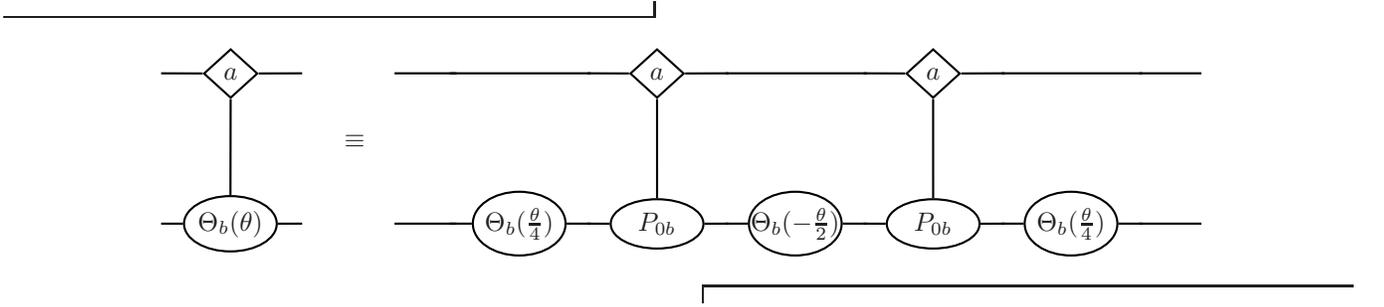
\begin{widetext}
  \begin{center}
    \kairocu{a}{\Shita{b}{\theta}}
    \kairoji{$\equiv$}
    \kairotsunagi\kairoichi{\Shita{b}{\frac{\theta}{4}}}%
    \kairocu{a}{P_{0b}}\kairoichi{\Shita{b}{-\frac{\theta}{2}}}%
    \kairocu{a}{P_{0b}}\kairoichi{\Shita{b}{\frac{\theta}{4}}}%
    \kairotsunagi
  \end{center}
\end{widetext}
Next we consider a single qudit operation:
\begin{equation*}
%  \label{eq:22}
  \Theta({\bm\theta})\ket{b}=
  \left\{
    \begin{array}{@{\,}rl}
      e^{-\kyo(\theta_1+\theta_2+\cdots+\theta_{d-1})}\ketrei,
      &\quad\text{if }\ket{b}=\ketrei\\[5pt]
      e^{\kyo\theta_b}\ket{b},&\quad\text{otherwise}
    \end{array}
  \right.,
\end{equation*}
where we abbreviate
\begin{equation*}
%  \label{eq:23}
  {\bm\theta}=(\theta_1,\theta_2,\ldots,\theta_{d-1}).
\end{equation*}
The matrix form is
\begin{equation*}
%  \label{eq:24}
  \Theta({\bm\theta})=\diag(e^{-\kyo(\theta_1+\theta_2+\cdots+\theta_{d-1})},
e^{\kyo\theta_1},\ldots,e^{\kyo\theta_{d-1}}).
\end{equation*}
Then the controlled operation
\begin{equation*}
  \tilde C^a(\Theta({\bm\theta}))\ket{c}\ket{b}=
  \left\{
    \setlength{\arraycolsep}{1pt}
    \begin{array}{@{\,}rl}
      \ket{a}\Theta({\bm\theta})\ket{b},&\quad\text{if }\ket{c}=\ket{a}\\[5pt]
      \ket{c}\ket{b},&\quad\text{otherwise}
    \end{array}
  \right.
\end{equation*}
is realized by
\begin{equation*}
  \tilde C^a(\Theta({\bm\theta}))=\prod^{d-1}_{b=1}\tilde C^a(\Shita{b}{\theta_b})
\end{equation*}
and the diagram is
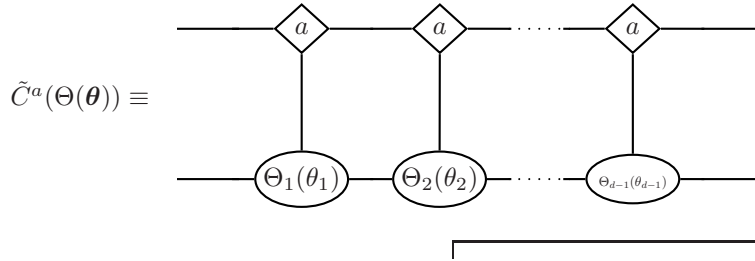
\begin{widetext}
  \begin{center}
    \kairoji{\hspace{-5em}$\tilde C^a(\Theta({\bm\theta}))\equiv$}\kairotsunagi%
    \kairocu{a}{\Shita{1}{\theta_1}}\kairocu{a}{\Shita{2}{\theta_2}}\kairoten%
    \kairocu{a}{\resizebox{6ex}{!}{$\Shita{d-1}{\theta_{d-1}}$}}\kairotsunagi
  \end{center}
\end{widetext}

\subsection{Construction of the controlled-$U$}
\label{sec:contr-contr-u}

First we construct the controlled-$U$ for not any unitary
transformation but special unitary transformation $W\in\mathrm{SU}(d)$.

For any $W\in\mathrm{SU}(d)$, there exists $V\in\mathrm{SU}(d)$ which satisfies
\begin{align*}
  W&=V^\dagger\Theta({\bm\theta})V,\\
  \Theta(\theta)&\equiv\diag(e^{-\kyo(\theta_1+\theta_2+\cdots+\theta_{d-1})},e^{\kyo\theta_1},e^{\kyo\theta_2},
  \cdots,e^{\kyo\theta_{d-1}}),
\end{align*}
where $\Theta(\theta)$ is an appropriate diagonal matrix.
By this fact we obtain the diagram of the controlled-$U$ for
$\mathrm{SU}(d)$ as follows:
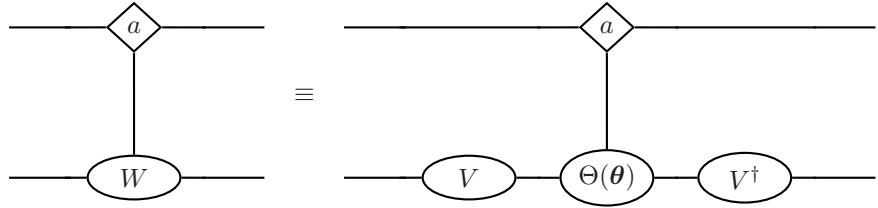
\begin{widetext}
  \begin{center}
    \kairotsunagi\kairocu{a}{W}\kairotsunagi\kairoji{$\equiv$}
    \kairotsunagi\kairoichi{V}\kairocu{a}{\Theta({\bm\theta})}%
    \kairoichi{V^\dagger}\kairotsunagi
  \end{center}
\end{widetext}

To extend the above result to $\mathrm{U}(d)$, we introduce the phase
gate $\tilde C^a(S)$:
\begin{equation*}
%  \label{eq:11}
  \tilde C^a(S)\ket{c}\ket{b}=
\left\{
  \begin{array}{rl}
    e^{\kyo\delta}\ket{a}\ket{b},&\quad\text{if }\ket{c}=\ket{a}\\[5pt]
    \ket{c}\ket{b},&\quad\text{otherwise}
  \end{array}
\right.
\end{equation*}
following \cite{9nin}.
In the similar way to qubit case, the diagram is given by
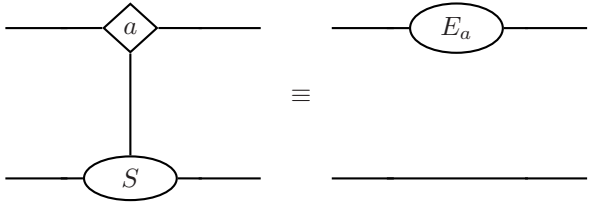
\begin{center}
  \kairotsunagi\kairocu{a}{S}\kairotsunagi\kairoji{$\equiv$}%
  \kairotsunagi\kaironi{E_a}\kairotsunagi
\end{center}
where
\begin{equation*}
%  \label{eq:25}
  E_a\equiv\diag(1,1,\ldots,1,\overset{\sacc{a\\\lor}}{e^{\kyo\delta}},1,\ldots,1)\in\mathrm{SU}(d).
\end{equation*}
Indeed, in two-qudit representation
\begin{equation*}
%  \label{eq:26}
  E_a\otimes\tani=\diag(\tani,\tani,\ldots,\tani,\overset{\sacc{a\\\lor}}{e^{\kyo\delta}}\tani,
  \tani,\ldots,\tani)
  =\tilde C^a(S).
\end{equation*}

Making use of the phase gate, we can construct the controlled-$U$
for $\mathrm{U}(d)$.
Any $U\in\mathrm{U}(d)$ is decomposed to
\begin{equation*}
  U=e^{\kyo\delta}W\qquad(W\in\mathrm{SU}(d)).
\end{equation*}
By this fact the controlled-$U$ for $U(d)$ is given by
\begin{equation*}
%  \label{eq:27}
  \tilde C^a(U)=\tilde C^a(W)\tilde C^a(S).
\end{equation*}
and the diagram is
\begin{center}
%  \kairotsunagi
\kairocu{a}{U}%
%\kairotsunagi
\kairoji{$\equiv$}%
%  \kairotsunagi
\kairocu{a}{W}\kaironi{E_a}%\kairotsunagi
\end{center}

\section{Construction of Unitary Transformation in two-Qudit}
\label{sec:constr-unit-transf}

In this section we construct unitary transformation
$\tilde U\in\mathrm{U}(d^2)$ in two-qudit making use of the controlled-$U$
in the previous section.
We follow the method by Deutch\cite{deutsch}.
In Sec.~\ref{sec:transf-basis-single} we show that any state is transformed 
to an arbitrary basis vector in a single qudit.
The similar result in two-qudit is shown in Sec.~\ref{sec:transf-basis-two}.
Then making use of this fact we construct unitary transformation in 
two-qudit in Sec.~\ref{sec:constr-unit-transf-1}.

\subsection{Transformation to a basis vector in a single qudit}
\label{sec:transf-basis-single}

A state in a single qudit is written by
\begin{equation*}
%  \label{eq:28}
  \ket{\fx}=\sum_{k=0}^{d-1}c_k\ket{k},\quad
  (c_k\in\fukuso,\ k=0,1,2,\ldots,d-1)
\end{equation*}
or, in the matrix form
\begin{equation*}
%  \label{eq:29}
  \fx=\tenchi{(c_0,c_1,\ldots,c_{d-1})}.
\end{equation*}

For any state $\ket{\fx}$, there exists a unitary transformation $U$
which satisfies
\begin{equation*}
%  \label{eq:30}
  N_0\ketrei=U\ket{\fx}
\end{equation*}
where
\begin{equation*}
%  \label{eq:31}
  N_n\equiv(\sum_{i=n}^{d-1}\vert c_i\vert^2)^{1/2}
\end{equation*}
or in the matrix form
\begin{equation*}
%  \label{eq:32}
  \tenchi{(N_0,0,\ldots,0)}=U\fx.
\end{equation*}
Such $U$ is concretely (but not necessarily efficiently) constructed
as follows.
We put
\begin{multline*}
%  \label{eq:33}
  h_k=
  \bordermatrix{
    &&&&{\footnotesize\mbox{$k$}}&{\footnotesize\mbox{$k+1$}}&&&\cr
    &1&&&&&&&\cr
    &&\ddots&&&&&&\cr
    &&&1&&&&&\cr
    {\footnotesize\mbox{$k$}}&&&&\frac{c_k}{N_k}&-\frac{N_{k+1}}{N_k}&&&\cr
    {\footnotesize\mbox{$k+1$}}&&&&\frac{N_{k+1}}{N_k}&-\frac{c_k^*}{N_k}&&&\cr
    &&&&&&1&&\cr
    &&&&&&&\ddots&\cr
    &&&&&&&&1\cr
  },
  \\
  (k=0,1,\ldots,d-3),
\end{multline*}
\begin{equation*}
%  \label{eq:34}
  h_{d-2}=\gyouretsu{
    1&&&&\\
    &\ddots&&&\\
    &&1&&\\
    &&&\frac{c_{d-2}}{N_{d-2}}&-\frac{c_{d-1}^*}{N_{d-2}}\\
    &&&\frac{c_{d-1}}{N_{d-2}}&-\frac{c_{d-2}^*}{N_{d-2}}
  },
\end{equation*}
then we obtain
\begin{equation*}
%  \label{eq:35}
  (h_{d-2}h_{d-3}\cdots h_1h_0)^\dagger\fx=\tenchi{(N_0,0,\ldots,0)}.
\end{equation*}
We also obtain an arbitrary basis vector as follows:
\begin{equation}
  \label{eq:36}
%  \tenchi{(0,0,,\ldots,\overset{k}{N_0},0,\ldots,0)}=U\fx.
  {}^t(0,0,,\ldots,\overset{k}{N_0},0,\ldots,0)=U\fx.
\end{equation}
Indeed, if we multiply the exchange operator after operating $h$'s, we obtain
\begin{multline*}
%  \label{eq:37}
  P_{0k}(h_{d-2}h_{d-3}\cdots h_1h_0)^\dagger\fx\\=P_{0k}{}^t(N_0,0,\ldots,0)\\
%  \tenchi{(0,0,,\ldots,\overset{k}{N_0},0,\ldots,0)}.
  ={}^t(0,0,,\ldots,\overset{k}{N_0},0,\ldots,0).
\end{multline*}
We temporarily call this operator as $T_k(\fx)$.
We note the operation again:
\begin{multline*}
%\begin{equation*}
%  \label{eq:38}
  T_k(\fx)\fx={}^t(0,0,,\ldots,\overset{k}{N_0},0,\ldots,0)
  \\(k=0,1,\ldots,d-1).
%\end{equation*}
\end{multline*}
%% The diagram is
%% \begin{center}
%%   \kairoaqudittsunagi\kairoaquditu{T_k(\fx)}\kairoaqudittsunagi
%% \end{center}

\subsection{Transformation to a basis vector in two-qudit}
\label{sec:transf-basis-two}

Making use of $T_k(\fx)$, we construct the unitary operator 
$\tilde S_b^a(\tilde\fx)$, which transforms a state $\tilde\fx$ in two-qudit
to a basis vector.

A state in two-qudit is written by
\begin{equation*}
%  \label{eq:39}
  \ket{\tilde\fx}=\sum_{i,j=0}^{d-1}c_{ij}\ket{i}\ket{j},
  \qquad(\sum_{i,j=0}^{d-1}\vert c_{ij}\vert^2=1),
\end{equation*}
or, in the matrix form
\begin{align*}
  \tilde\fx&=
  {}^t(c_{00},c_{01},\ldots,c_{0,d-1},c_{10},c_{11},\ldots,\\&\hspace{4em}c_{1,d-1},\ldots
  \ldots,c_{d-1,0},\ldots,c_{d-1,d-1})\\
  &=\tenchi{(\fx_0,\fx_1,\ldots,\fx_{d-1})}
\end{align*}
where
\begin{equation*}
  \phantom{=}\fx_i\equiv(c_{i0},c_{i1},\ldots,c_{i,d-1})\qquad(i=0,1,\ldots,d-1).
\end{equation*}
Then the diagram is
\begin{widetext}
\begin{center}
\kairohako{\tilde S_b^a(\tilde\fx)}
\kairoji{$\equiv$} % \kairotsunagi%
%\kairocu{\scriptsize\mbox{$d-1$}}{T_b(\fx_{d-1})}\kairotsunagi%
\kairocu{\resizebox{!}{1ex}{$d-1$}}{\resizebox{3em}{!}{$T_b(\fx_{d-1})$}}%
%\kairotsunagi%
%\kairocu{\scriptsize\mbox{$d-2$}}{T_b(\fx_{d-2})}%
\kairocu{\resizebox{!}{1ex}{$d-2$}}{\resizebox{3em}{!}{$T_b(\fx_{d-2})$}}%
\kairotsunagi%
\kairoten\kairocu{0}{T_b(\fx_{0})}\kairocuni{b}{T_a(\fy)}
\end{center}
\end{widetext}
where we put
\begin{equation*}
%  \label{eq:40}
  \fy\equiv(\|\fx_0\|,\|\fx_1\|,\ldots,\|\fx_{d-1}\|)\in V_d(\fukuso).
\end{equation*}
In this circuit we find
\begin{equation*}
%  \label{eq:41}
  \tilde S^a_b(\tilde\fx)\ket{\tilde\fx}=\ket{a}\ket{b}.
\end{equation*}

\subsection{Construction of unitary transformation in two-qudit}
\label{sec:constr-unit-transf-1}

In this subsection, finally, we construct unitary
transformation in two-qudit.
Let $\tilde U\in\mathrm{U}(d^2)$ be the unitary transformation
whose eigenvalues are $e^{\kyo\sigma(a,b)}$ ($a,b=0,1,2,\ldots,d-1$)
and the corresponding eigenstates $\ket{\sigma(a,b)}$.

We introduce $\tilde X(a,b)$ whose operation is
\begin{equation*}
%  \label{eq:42}
  \tilde X(a,b)\ket{c}\ket{d}=
  \left\{
    \begin{array}{@{\,}rl}
      e^{\kyo\kochiab}\ket{a}\ket{b},
      &\text{if }\ket{c}\ket{d}=\ket{a}\ket{b}\\[5pt]
      \ket{c}\ket{d},&\text{otherwise}
    \end{array}
  \right..
\end{equation*}
$\tilde X(a,b)$ is given by
\begin{equation*}
%  \label{eq:44}
  \tilde X(a,b)=\tilde C^a(X(a,b))
\end{equation*}
where $X(a,b)$ is a single qudit gate:
\begin{equation*}
%  \label{eq:42}
  \tilde X(a,b)\ket{c}\ket{d}=
  \left\{
    \begin{array}{@{\,}rl}
      e^{\kyo\kochiab}\ket{a}\ket{b},
      &\text{if }\ket{c}\ket{d}=\ket{a}\ket{b}\\[5pt]
      \ket{c}\ket{d},&\text{otherwise}
    \end{array}
  \right..
\end{equation*}
The diagram is
\begin{center}
  \kairotsunagi\kairohako{\tilde X(a,b)}\kairotsunagi\kairoji{$\equiv$}%
  \kairotsunagi\kairocu{a}{X(a,b)}\kairotsunagi
\end{center}
In the matrix form
%\begin{equation*}
\begin{multline*}
%  \label{eq:45}
  \tilde X(a,b)\equiv\diag(\tani,\tani,\ldots,\tani,X(a,b),\tani)\\
  =\bordermatrix{
    &&&&da+b&&&\cr
    &1&&&&&&\cr
    &&\ddots&&&&&\cr
    &&&1&&&&&\cr
    da+b&&&&e^{\kyo\kochiab}&&&\cr
    &&&&&1&&\cr
    &&&&&&\ddots&\cr
    &&&&&&&1\cr
  }
%\end{equation*}
\end{multline*}
with
%\begin{equation*}
\begin{multline*}
%  \label{eq:46}
  X(a,b)\equiv\diag(1,1,\ldots,1,e^{\kyo\kochiab},1,\ldots,1)\\
  =\bordermatrix{
    &&&&b&&&\cr
    &1&&&&&&\cr
    &&\ddots&&&&&\cr
    &&&1&&&&&\cr
    b&&&&e^{\kyo\kochiab}&&&\cr
    &&&&&1&&\cr
    &&&&&&\ddots&\cr
    &&&&&&&1\cr
  }.
%\end{equation*}
\end{multline*}
or, in component,
\begin{align*}
%  \label{eq:47}
  (\tilde X(a,b))_{ij}&=\delta_{ij}\{1+\delta_{i,da+b}(-1+e^{\kyo\kochiab})\},\\
  (X(a,b))_{ij}&=\delta_{ij}\{1+\delta_{ib}(-1+e^{\kyo\kochiab})\}.
\end{align*}

By the result of Sec.~\ref{sec:transf-basis-two}, there exists
$\tilde S^a_b(\kochiab)$ which satisfies
\begin{equation*}
%  \label{eq:48}
  \tilde S^a_b(\kochiab)\ket{\kochiab}=\ket{a}\ket{b},\quad
  (a,b=0,1,\ldots,d-1).
\end{equation*}
Then we introduce the operator
\begin{equation*}
%  \label{eq:49}
  \tilde Z(a,b)\equiv\tilde S^{-1}(\kochiab)\tilde X(a,b)\tilde S(\kochiab).
\end{equation*}
The diagram is
\begin{widetext}
\begin{center}
  \kairotsunagi\kairohako{\tilde Z(a,b)}\kairotsunagi\kairoji{$\equiv$}%
  \kairotsunagi\kairohako{\rotatebox{270}{$\tilde S^{-1}(\kochiab)$}}%
  \kairohako{\tilde X(a,b)}%
  \kairohako{\rotatebox{270}{$\tilde S(\kochiab)$}}\kairotsunagi
\end{center}
\end{widetext}
They satisfy
\begin{equation*}
  \tilde Z(a,b)\ket{\kochi{c}{d}}=
  \left\{
    \begin{array}{@{\,}rl}
      e^{\kyo\kochiab}\ket{\kochiab},&\ket{\kochi{c}{d}}=\ket{\kochiab}\\[5pt]
      \ket{\kochi{c}{d}},&\ket{\kochi{c}{d}}\neq\ket{\kochiab}
    \end{array}
  \right..
\end{equation*}

Finally we construct $\tilde U$ by
\begin{equation*}
%  \label{eq:50}
  \tilde U=\prod_{a,b=0}^{d-1}\tilde Z(a,b).
\end{equation*}
The diagram is
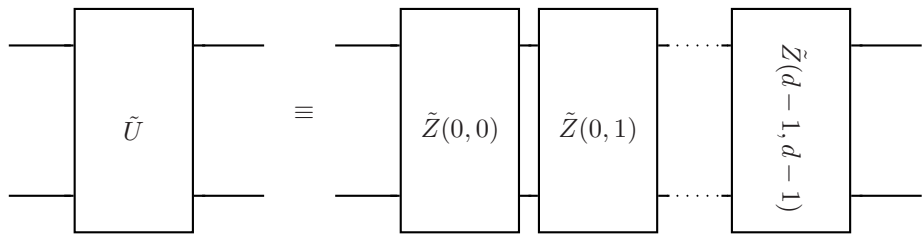
\begin{widetext}
\begin{center}
  \kairotsunagi\kairohako{\tilde U}\kairotsunagi\kairoji{$\equiv$}%
  \kairotsunagi\kairohako{\tilde Z(0,0)}%
  \kairohako{\tilde Z(0,1)}\kairoten%
  \kairohako{\rotatebox{270}{$\tilde Z(d-1,d-1)$}}\kairotsunagi
\end{center}
\end{widetext}
Indeed, this operator satisfies
\begin{equation*}
%%   (\prod_{a,b=0}^{d-1}\tilde Z(a,b))\ket{\kochiab}=e^{\kyo\kochiab}\ket{\kochiab},\quad
%%   (a,b=0,1,2,\ldots,d-1),
  \tilde U\ket{\kochiab}=e^{\kyo\kochiab}\ket{\kochiab},\quad
  (a,b=0,1,2,\ldots,d-1),
\end{equation*}
to be diagonal in the eigenstate of $\tilde U$.
Thus we find that this is the circuit which perform $\tilde U$.

\section{Discussions}
\label{sec:discussions}

We have constructed the controlled-$U$ gate in qudit and unitary 
transformation in two-qudit.
In the similar way to extend the controlled-$U$ to the $\text{controlled}^n$
-$U$ in qubit\cite{9nin}, the controlled-$U$ will be extended in qudit.

However, in this method as well as in the qubit case, the larger $n$ is 
the larger steps exponentially and the more difficult calculation is.
To avoid the problem some quite new ideas may be required.

When we construct the controlled-$U$, we do not use the Euler
decomposition but use the diagonalization.
The former may fit the property of laser and have advantage to 
construct the circuit with laser operation.
However, when number of states in one qudit is large, the decomposition
seems to be complicated~\cite{tilma1,tilma2}.
Our method may not fit the property of laser, however, the method of 
diagonalization is well-known and even when number of states is large,
network is relatively easy built.
Laser is not necessarily needed to construct networks
and there may exist physical systems fit to the diagonalization.

We adopt the controlled-$M_b$ gate as the elementary gate of
the controlled operation.
This choice stems from the notion that in physical systems
manipulation between only two states is allowed.
However, there may exist physical systems in which more than two states
are manipulated at once.
In such cases, we can choose other gate to decrease steps of calculation.

As stated in the introduction, there are the problem of decoherence
through interaction with environment in construction of networks.
Taking this fact into consideration, qudit has advantage to qubit.
However, to realize qudit in physical systems, for example,
if we make use of energy levels of electron in an atom,
the energy differences between high excited states are very small, 
thus in actual not so many levels are used in construction.

%% \begin{center}
%%   \large\bfseries Acknowledgments
%% \end{center}
\begin{acknowledgments}
  We thank K. Fujii for important ideas and useful discussions.
\end{acknowledgments}

\end{document}